\documentclass[conference]{IEEEtran}
\IEEEoverridecommandlockouts
\usepackage{cite}
\usepackage{amsmath,amssymb,amsfonts}
\usepackage{algorithmic}
\usepackage{graphicx}
\usepackage{textcomp}
\usepackage{xcolor}
\def\BibTeX{{\rm B\kern-.05em{\sc i\kern-.025em b}\kern-.08em
    T\kern-.1667em\lower.7ex\hbox{E}\kern-.125emX}}
\begin{document}

\title{Spectrogram Feature Losses for Music Source Separation}

\author{\IEEEauthorblockN{Abhimanyu Sahai}
\IEEEauthorblockA{\textit{ETH Z{\"u}rich} \\
asahai@ethz.ch}
\and
\IEEEauthorblockN{Romann Weber}
\IEEEauthorblockA{\textit{Disney Research Z{\"u}rich} \\
romann.weber@disneyresearch.com}
\and
\IEEEauthorblockN{Brian McWilliams}
\IEEEauthorblockA{\textit{Disney Research Z{\"u}rich} \\
brian.mcwilliams@disneyresearch.com}
}

\maketitle

\begin{abstract}
In this paper we study deep learning-based music source separation, and explore using an alternative loss to the standard spectrogram pixel-level L2 loss for model training. Our main contribution is in demonstrating that adding a high-level feature loss term, extracted from the spectrograms using a VGG net, can improve separation quality vis-a-vis a pure pixel-level loss. We show this improvement in the context of the MMDenseNet, a State-of-the-Art deep learning model for this task, for the extraction of drums and vocal sounds from songs in the \textit{musdb18} database, covering a broad range of western music genres. We believe that this finding can be generalized and applied to broader machine learning-based systems in the audio domain.
\end{abstract}

\section{Introduction}

Music source separation is a problem that has been studied for a few decades now: given an audio track with several instruments mixed together (a regular MP3 file, for example), how can it be separated into its component instruments? The obvious application of this problem is in music production - creating karaoke tracks, highlighting select instruments in an audio playback, etc. There is another reason why this is a useful problem to study: it acts as a powerful enabler for several other applications in music informatics. This is because complex, multi-instrument music tracks are not easily processed by such algorithms in their raw audio form. However, once individual instruments have been isolated from such a track, they can relatively easily be transcribed by contemporary algorithms.

Up until the early 2010s, the most common approaches to this problem were not data-driven, but rooted in exploiting known statistical properties of music signals, or in signal processing theory. However, as with many fields, that has changed in the last few years with the advent of cheaper computing power and proliferation of research in machine learning. The best performance on this problem is currently achieved by deep learning-based methods. These methods feed the mixture at the input of the network, and the source(s) as targets (or rather typically the spectrograms of the input/output, since many of the patterns to be discovered are in the frequency domain) to learn a function mapping between the two.

These deep learning approaches use a pixel-level loss as the cost function, averaging the L2 losses between corresponding pixels in the output and target spectrograms. (The term 'pixel' here, and in the rest of the paper is used to denote time-frequency bins in the spectrogram, because the spectrogram is treated as an image for the purpose of our work.) However, we believe that this is not the ideal loss function for this problem. This is because it does not explicitly give weight to higher-level patterns in spectrograms, which could exhibit similarity between similar pieces of audio. For example, non-pitched instruments like drums have signal present across frequencies, and therefore exhibit vertical lines in their spectrograms. On the other hand, vocal spectrograms display harmonicity, i.e. horizontal lines. Thus, we propose that pairing the pixel-level L2 loss with a loss between higher-level patterns extracted from the spectrogram could lead to improved performance. For the latter, we port the loss terms developed by the authors in \cite{b1} for the visual domain, treating spectrograms as images for this purpose. This is not an ideal treatment, and better alternatives will be discussed in Section \ref{future-work} on future work.

The rest of this paper is organized as follows: In the following sections we introduce the core deep learning model for music source separation that we have utilized in our work, and briefly summarize the learning from \cite{b1} in using VGG feature maps to compute the spectrogram feature losses. After laying down related work, we describe in detail our experiments and their results. We summarize the implications of these results and finally discuss ideas to build further on this work. 

\section{Related Work}

To the best of our knowledge, there is no existing work on the application of such spectrogram feature losses to music source separation. The general idea of applying feature/style reconstruction losses as proposed in \cite{b1} for the visual domain, to an audio domain problem has been explored by some researchers, with mixed results. In \cite{b8}, the authors propose an audio style transfer using, as one of the approaches, style reconstruction losses extracted using the VGG network, similar to \cite{b1}. In their case, the VGG does not yield results of acceptable quality (as per subjective tests) but using a shallow CNN does. In \cite{b9}, the authors explore audio generation as an audio style transfer problem, using similar loss terms. More generally, the idea of perceptual losses for audio is still an open area of research, where the task is to find loss measures that correlate better with subjective measures of audio quality. However, while the feature losses we explore in our work are derived from perceptual losses in the image domain, they are more directly a descriptor of visual patterns in audio spectrograms than being a perceptual descriptor of the underlying audio.

\section{MMDenseNets for Music Source Separation}

Multi-scale Multi-band DenseNets, or MMDenseNets are a CNN-based deep network model for music source separation. They were proposed in \cite{b2}, and variations of this model achieve the current State-of-the-Art performance on the music source separation task, based on the SiSEC - the Signal Separation Evaluation Campaign. This is a benchmark competition for this task that we discuss in greater detail in Section \ref{dbm}. In this section, we provide a brief overview of this model. 

At the input of the MMDenseNet is the spectrogram of the mixed-up song, in its STFT (Short-Time Fourier Transform) representation. Each source to be separated has its own network and set of weights, and for each network, the training targets comprise the corresponding pure source spectrograms. Since this is a real-valued neural network, the phase of the mixture spectrogram is isolated and only the magnitude is fed into the network. Similarly, during training, the target consists only of the magnitude of the source spectrogram. In order to recover the estimated time-domain source signal during inference, the phase of the input mixture spectrogram is directly applied to each source spectrogram, and an inverse-STFT taken of the result. In case the data is stereophonic, i.e. contains more than one channel, this information is fed into the MMDenseNet as a multi-channel spectrogram image.

\begin{figure}
\begin{center}
\includegraphics[scale=0.15]{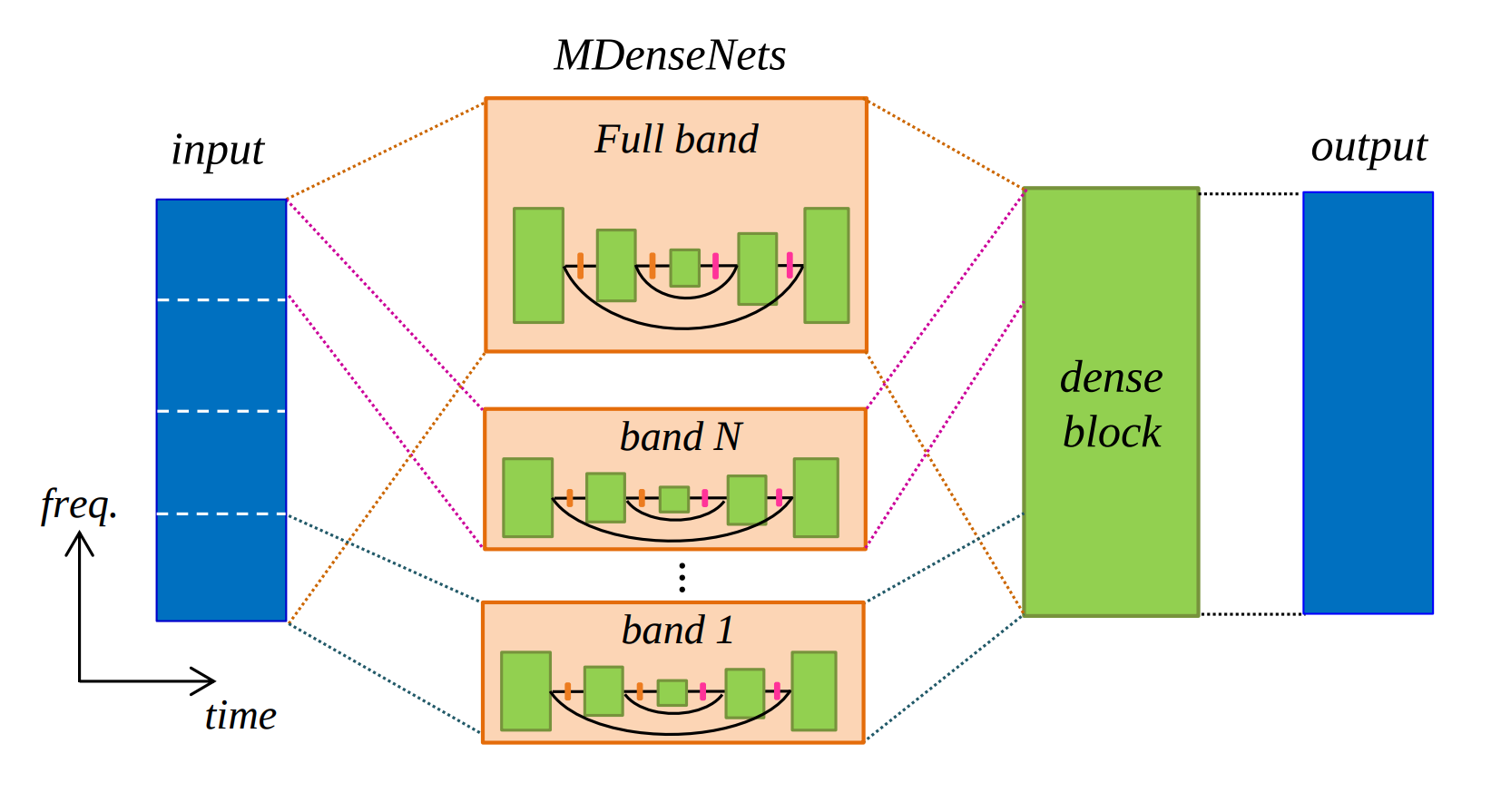}
\end{center}
\caption{Illustration of complete MMDenseNet architecture. Reproduced with permission from
\cite{b2}\label{fig:Illustration-of-complete-mmdensenet}}
\end{figure}

The network architecture itself is based on the DenseNet, which is a deep CNN where every layer's output is directly fed to every other layer succeeding it. For greater detail on the DenseNet architecture, the reader is referred to the original paper \cite{b3}. Furthermore, while the original DenseNet is a classifier and periodically downsamples the original image, in the current application an image needs to be created at the output. For this purpose, the MMDenseNet includes an upsampling path, also comprised of DenseNet blocks, resulting in an autoencoder style architecture. What makes the MMDenseNet especially unique is its use of sub-band networks - in simple language, rather than sharing the convolution kernel across the spectrogram image, it trains separate convolutional layers (and therefore kernels) for different frequency bands. It achieves this in practice by splitting the input spectrogram along the frequency axis into two or more sub-images - each of which can be thought of as representing a sub-(frequency)band image. Each of these sub-band images is propagated through its own DenseNet autoencoder as described above. Towards the output, feature maps from these sub-band DenseNets are joined back along the frequency axis. The MMDenseNet architecture is illustrated in Figure \ref{fig:Illustration-of-complete-mmdensenet}.

As a post-processing step during inference, the predictions of the network for each source are scaled, for each time-frequency bin, so that their sum is equal to the original mixture at the corresponding time-frequency bin. This is akin to single-channel Wiener filtering, and is also part of the procedure established in \cite{b2}.

\section{Spectrogram Feature Loss}\label{vgg-loss}

In this section, we explain how a high-level spectrogram feature loss can be computed using the VGG network. This network refers to a deep convolutional neural network developed by Oxford's Visual Geometry Group (VGG - hence the name of the network) for the purpose of image classification \cite{b4}. The network uses a succession of convolution, \textit{relu} activation and max-pooling layers to extract image features, plugging in a fully connected layer followed by a softmax layer in the end for performing the classification task. This network was among the winners in the ImageNet challenge in 2014.

While the purpose of the VGG network as it was developed was image classification, it is of interest to us for our problem because it can also be viewed as a feature extractor. Successive hidden layers of the network compute ’higher-level’ image features, like shapes and forms. So, instead of comparing two images only on their pixel values, we could use the VGG as a feature extractor to obtain high-level features and compare the images on these features as well. It was this insight that was used by the authors in \cite{b1} to do a style-transfer between two images.

For the purpose of this work, we treat the high-level spectrogram feature loss calculation as a black-box, computed exactly as in \cite{b1}, i.e. using the VGG network and computing two related loss terms - the feature and style reconstruction losses. We use the same layers of the VGG network for computing these loss terms as in \cite{b1}. Throughout our experiments, which we shall describe in Section \ref{experiments}, we give a weight of 0.5 to the regular pixel-level L2 loss and 0.25 to each of these two high-level feature losses. In the rest of this paper, we use the term \textit{composite spectrogram loss} for the weighted combination. We arrived at these values for the weights empirically. In particular, we also tried using only the high-level feature losses but found the performance to be inferior for this setting. Ideally, we would use an analog to the VGG network for the audio or music domain, to optimize for extracting audio-specific features. However no such publicly available and rigorously tested network exists. In Section \ref{future-work} on future work, we discuss how this black-box calculation can be better customized for this application.

\section{Experiments}

\subsection{Dataset, Benchmarks and Metrics}\label{dbm}

The SiSEC is a biennial forum where researchers in signal separation - across a variety of signal domains (eg. bio-medical, music, etc.) compare the performance of their algorithms on a standardized task. The music source separation sub-task currently involves separation of 50 professionally recorded stereo tracks, across varying genres like pop, rock, rap etc., into \textit{vocals}, \textit{drums}, \textit{bass} and \textit{other}, i.e. the collection of remaining instruments as one track. Since the researchers report detailed standardized metrics, and also discuss their approach at varying lengths, this is a good resource to glean the State-of-the-Art for this problem.

For this sub-task SiSEC provides a dataset called \textit{musdb18} \cite{b10}. It consists of 150 professionally-recorded tracks across genres, of which the actual testing is to be done on 50 tracks, while the other 100 can be used for training in supervised approaches. For each track, the ’true’ isolated \textit{vocals}, \textit{drums}, \textit{bass} and \textit{other} tracks are provided, along with the main mixed track.

Performance is evaluated on a collection of specialized metrics developed and widely used by the research community in blind source separation, called BSS Eval \cite{b5}. These measures are somewhat akin to an SNR measure. In the following sections of this paper, we will compare performances on the Signal-to-Distortion Ratio (SDR) as it is the overarching metric that encompasses the other metrics.

\subsection{Baseline Model Implementation}\label{baseline-model}

Since the MMDenseNet model is not open-sourced, we created our own implementation following the general guidelines listed in \cite{b2} and applied it to the SiSEC 2018 task. The parameters for the MMDenseNet architecture in our implementation are the same as those given in Table 1 of \cite{b2}. Other important implementation details are as follows: We use 2048 samples for the FFT, with a hop rate of 1024. Each spectrogram contains 128 time frames. We use RMSProp for optimization, starting with a learning rate of 0.001 and dropping it to 0.0001 when learning saturates. Finally, we use a bottleneck-compressed version of the DenseNet as explained in \cite{b4}, with a factor of 4 for the bottleneck and a factor of 0.2 for the compression.  

\begin{figure}
\begin{center}
\includegraphics[scale=0.4]{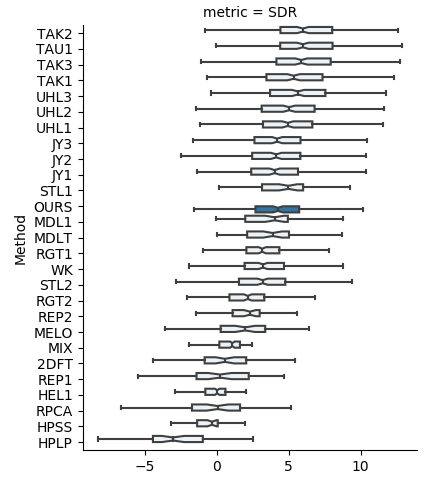}
\end{center}

\caption{Boxplots (over all the test songs) of our baseline model's performance compared to other SiSEC 2018 submissions,
for the \textit{vocals} source. The SDR should be viewed as the overall summary metric, with a higher SDR implying better performance. \label{fig:MMDenseNet-vocals}}
\end{figure}

As described in the SiSEC 2018 paper \cite{b6}, we calculate the median value of the SDR for each source over all time windows. Figure \ref{fig:MMDenseNet-vocals} shows the boxplot of the SDR thus obtained over all songs in the \textit{musdb18} test database, for each method submitted to SiSEC 2018, for the \textit{vocals} source as an illustration of our relative performance. Our relative performance is similar for the other sources. Our method is labelled OURS. While our focus was more to get a reasonably performant working implementation of a deep learning music source separation system to be able to compare the pixel-level loss with composite spectrogram losses, we do come close to the State-of-the-Art as well. It should be noted that among the submissions in Figure \ref{fig:MMDenseNet-vocals}, TAK1, TAK2 and TAK3 are based on the MMDenseNet. The gap in performance between our model and these submissions can be explained by a mix of reasons - chiefly, their use of data augmentation, the use of an LSTM layer in addition to the DenseNet CNNs, and the use of specialized architectures for different sources (For eg., increasing complexity of the lower frequency sub-band for the \textit{bass} source).

\subsection{Pixel-level vs. Composite Spectrogram Loss Comparison Methodology}\label{experiments} 

With the baseline model implemented as above, we conducted a series of experiments to compare its performance with pixel loss, with the same model tuned with the composite spectrogram loss as defined in Section \ref{vgg-loss}. Below we describe the settings for each experiment. In all the experiments, training was done with the development set of the \textit{musdb18} database and the reported SDR is on its test set. Our experiments cover the sources \textit{vocals}, \textit{drums} and \textit{bass}.
 
\begin{itemize}
\item \textbf{Experiment A:} In this experiment, we compared the performance of the \textit{vocals} source isolation obtained by the pixel loss-tuned model with that of the composite spectrogram loss-tuned model. The parameter settings of the model in both the cases were identical and the same as those described in Section \ref{baseline-model}. We repeated this experiment four times, to reduce false inferences due to experimental randomness and thus to be able to comment on the statistical significance of the observed difference in performance between the two models, if any. Machine learning optimization is a random process - with some of the randomness introduced by the optimization algorithm, and some introduced by the parallel computing typically used for the optimization (eg. GPUs). We used Keras as our implementation framework, and while it can control for the former source of randomness through the use of random seeds, there is currently no way to control for the latter.
\item \textbf{Experiment B:} This was same as the above experiment, conducted for the \textit{drums} source (instead of \textit{vocals}).
\item \textbf{Experiment C:} This was also same as the above experiment, conducted for the \textit{bass} source.
\item \textbf{Experiment D:} In this experiment, we once again compared the \textit{vocals} source. However we did this with a single-channel model in place of the stereo (two-channel) model used in the above experiments (The \textit{musdb18} songs are available as two-channel recordings. A single-channel version can be created by averaging the two channels). The motivation to do this experiment was to test the composite spectrogram loss under more diverse use-cases and settings. Like the above experiments, this was conducted four times as well.
\end{itemize}

\section{Results and Discussion}

We discuss the results for each of the experiments described in the previous section.

\begin{itemize}

\item \textbf{Experiment A:} In Table \ref{tab:experiment-a-results}, we display:
\begin{enumerate}
\item The pixel-level L2 loss value obtained for the validation set upon convergence for both the models, for the \textit{vocals} source, in four independent runs. It should be noted, for the composite spectrogram loss-tuned model, that the pixel-level L2 loss is one of the components of the overall loss, as explained in Section \ref{vgg-loss}. For this model, we chose the epoch with the minimum composite validation loss, as one would usually do, but report in this table only the pixel-level L2 loss component, for a like-to-like comparison.
\item The SDR value obtained over the \textit{musdb18} test dataset by both the models, for the \textit{vocals} source, in the above four independent runs. The figure reported here is the median over the test dataset, as explained in Section \ref{baseline-model}.
\end{enumerate}
While there seems to be a visible difference in performance between the two models, with the composite spectrogram loss-tuned model outperforming the pixel loss-tuned model, we run the SDR results through a t-test for statistical rigor. The output from these tests conducted in R is also displayed in Table \ref{tab:experiment-a-results}. The differences are significant at a 5\% significance level. On this sample, the composite spectrogram loss-tuned model delivers a 0.27 dB improvement in performance.

\begin{table}
\caption{Comparison of source separation performance for the \textit{vocals} source between the pixel loss-tuned model (Model 1) and the composite spectrogram loss-tuned model (Model 2). Lower val. loss and higher SDR are better}
\begin{center}
\begin{tabular}{|c|c|c|c|c|}
\hline 
Run & \multicolumn{2}{c|}{Min. pixel val. loss (L2)} & \multicolumn{2}{c|}{SDR (dB)}\tabularnewline
\hline 
\hline 
 & Model 1 & Model 2 & Model 1 & Model 2\tabularnewline
\hline 
1 & $0.59$ & $\textbf{0.47}$ & $3.70$ & $\textbf{3.98}$\tabularnewline
\hline 
2 & $0.59$ & $\textbf{0.50}$ & $3.72$ & $\textbf{3.93}$\tabularnewline
\hline 
3 & $0.59$ & $\textbf{0.50}$ & $3.83$ & $\textbf{4.06}$\tabularnewline
\hline 
4 & $0.60$ & $\textbf{0.48}$ & $3.73$ & $\textbf{3.84}$\tabularnewline
\hline
\end{tabular}
\end{center}
\begin{center}
\bigskip
Welch Two Sample t-test
\begin{tabular}{lc}
\hline 
t-statistic & -4.52\tabularnewline
df & 4.47\tabularnewline
p-value & 0.008\tabularnewline
Mean SDR with pixel loss & 4.32\tabularnewline
Mean SDR with composite spectrogram loss & 4.59\tabularnewline
95\% confidence interval (Difference of means) &  -0.43, -0.11\tabularnewline
\hline 
\end{tabular}
\end{center}
\label{tab:experiment-a-results}
\end{table}

\item \textbf{Experiment B:} Similar to the above experiment, Table \ref{tab:experiment-b-results} shows the validation pixel-level L2 loss upon convergence, and the median SDR obtained over the \textit{musdb18} test set for both the models, for the source \textit{drums}. The table also gives the results of the t-test to check if the SDR results are significantly different. We can see, once again, that the composite spectrogram loss-tuned model outperforms the pixel loss-tuned model. However, the 5\% significance is more borderline for \textit{drums}. On this sample, the VGG loss model delivers a 0.18 dB improvement in performance.

\begin{table}
\caption{Comparison of source separation performance for the \textit{drums} source between the pixel loss-tuned model (Model 1) and the composite spectrogram loss-tuned model (Model 2). Lower val. loss and higher SDR are better}
\begin{center}
\begin{tabular}{|c|c|c|c|c|}
\hline 
Run & \multicolumn{2}{c|}{Min. pixel val. loss (L2)} & \multicolumn{2}{c|}{SDR (dB)}\tabularnewline
\hline 
\hline 
 & Model 1 & Model 2 & Model 1 & Model 2\tabularnewline
\hline 
1 & $0.46$ & $\textbf{0.37}$ & $4.70$ & $\textbf{4.88}$\tabularnewline
\hline 
2 & $0.46$ & $\textbf{0.37}$ & $4.53$ & $\textbf{4.65}$\tabularnewline
\hline 
3 & $0.48$ & $\textbf{0.37}$ & $4.52$ & $\textbf{4.71}$\tabularnewline
\hline 
4 & $0.46$ & $\textbf{0.38}$ & $4.64$ & $\textbf{4.88}$\tabularnewline
\hline 
\end{tabular}
\end{center}
\begin{center}
\bigskip
Welch Two Sample t-test
\begin{tabular}{lc}
\hline 
t-statistic & -2.49\tabularnewline
df & 5.53\tabularnewline
p-value & 0.051\tabularnewline
Mean SDR with pixel loss & 4.60\tabularnewline
Mean SDR with composite spectrogram loss & 4.78\tabularnewline
95\% confidence interval (Difference of means) &  -0.37, 0.00\tabularnewline
\hline 
\end{tabular}
\end{center}
\label{tab:experiment-b-results}
\end{table}

\item \textbf{Experiment C:} Similar to the above experiment, Table \ref{tab:experiment-c-results} shows the validation pixel-level L2 loss upon convergence, and the median SDR obtained over the \textit{musdb18} test set for both the models, for the source \textit{bass}. While the composite spectrogram loss-tuned model consistently converges to a lower validation L2 loss, in terms of SDR performance the two models seem to be nearly identical, at least based on these samples. We do not run these SDRs through a t-test.

\begin{table}
\caption{Comparison of source separation performance for the \textit{bass} source between the pixel loss-tuned model (Model 1) and the composite spectrogram loss-tuned model (Model 2). Lower val. loss and higher SDR are better}
\begin{center}
\begin{tabular}{|c|c|c|c|c|}
\hline 
Run & \multicolumn{2}{c|}{Min. pixel val. loss (L2)} & \multicolumn{2}{c|}{SDR (dB)}\tabularnewline
\hline 
\hline 
 & Model 1 & Model 2 & Model 1 & Model 2\tabularnewline
\hline 
1 & $0.64$ & $\textbf{0.49}$ & $\textbf{4.10}$ & $4.03$\tabularnewline
\hline 
2 & $0.61$ & $\textbf{0.50}$ & $4.00$ & $\textbf{4.01}$\tabularnewline
\hline 
3 & $0.61$ & $\textbf{0.51}$ & $4.09$ & $\textbf{4.14}$\tabularnewline
\hline 
4 & $0.63$ & $\textbf{0.51}$ & $4.04$ & $\textbf{4.06}$\tabularnewline
\hline 
\end{tabular}
\end{center}
\label{tab:experiment-c-results}
\end{table}

\item \textbf{Experiment D:} Table \ref{tab:experiment-d-results} shows the validation pixel-level L2 loss upon convergence, and the median SDR obtained over the \textit{musdb18} test set for both the models, for the source \textit{vocals}, for a single-channel model. The table also gives the results of the t-test to check if the SDR results are significantly different. We can see that the composite spectrogram loss-tuned model outperforms the pixel loss-tuned model at a 5\% significance level for the \textit{vocals} source under these settings as well. On this sample, the former delivers a 0.2 dB improvement in performance.

\begin{table}
\caption{Comparison of source separation performance for the \textit{vocals} source for a single-channel model between the pixel loss-tuned model (Model 1) and the composite spectrogram loss-tuned model (Model 2). Lower val. loss and higher SDR are better}
\begin{center}
\begin{tabular}{|c|c|c|c|c|}
\hline 
Run & \multicolumn{2}{c|}{Min. pixel val. loss (L2)} & \multicolumn{2}{c|}{SDR (dB)}\tabularnewline
\hline 
\hline 
 & Model 1 & Model 2 & Model 1 & Model 2\tabularnewline
\hline 
1 & $0.59$ & $\textbf{0.47}$ & $3.70$ & $\textbf{3.98}$\tabularnewline
\hline 
2 & $0.59$ & $\textbf{0.50}$ & $3.72$ & $\textbf{3.93}$\tabularnewline
\hline 
3 & $0.59$ & $\textbf{0.50}$ & $3.83$ & $\textbf{4.06}$\tabularnewline
\hline 
4 & $0.60$ & $\textbf{0.48}$ & $3.73$ & $\textbf{3.84}$\tabularnewline
\hline 
\end{tabular}
\end{center}
\begin{center}
\bigskip
Welch Two Sample t-test
\begin{tabular}{lc}
\hline 
t-statistic & -3.81\tabularnewline
df & 5.06\tabularnewline
p-value & 0.012\tabularnewline
Mean SDR with pixel loss & 3.75\tabularnewline
Mean SDR with composite spectrogram loss & 3.95\tabularnewline
95\% confidence interval (Difference of means) &  -0.35, -0.07\tabularnewline
\hline 
\end{tabular}
\end{center}
\label{tab:experiment-d-results}
\end{table}	

\end{itemize}

\begin{figure}
\begin{center}
\includegraphics[scale=0.35]{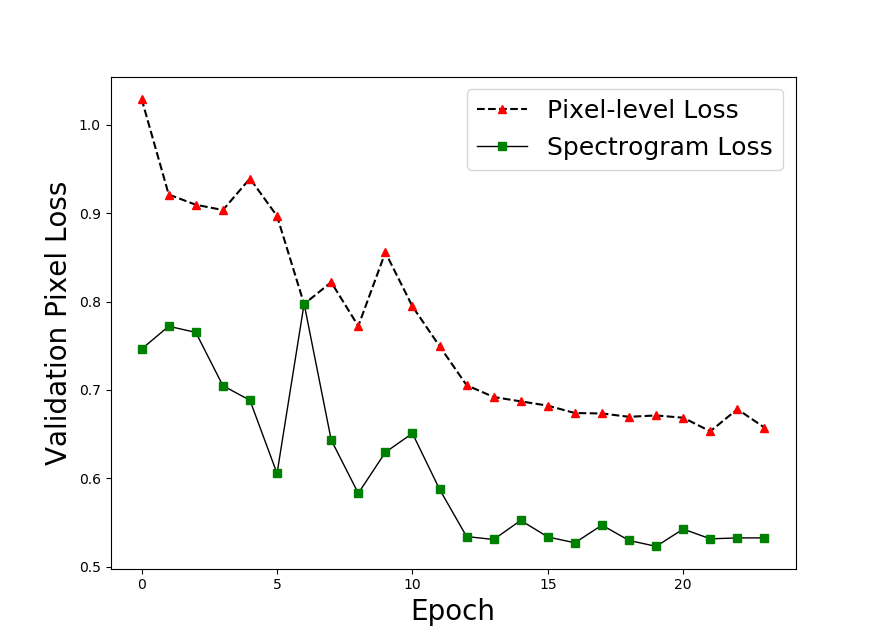}
\end{center}
\caption{Trajectory of validation pixel-level L2 losses for the composite spectrogram loss-tuned model vs. pixel-level loss-tuned model when training for $24$ epochs for the \textit{vocals} source \label{fig:L2-loss-trajectories}}
\end{figure}

We also show, in Figure \ref{fig:L2-loss-trajectories}, the validation pixel-level L2 loss trajectory for both the models, averaged across the four runs, for Experiment A (two-channel vocals). The trajectories for the other sources are similar. The difference in performance between the two models is once again evident from this plot, with the composite spectrogram loss-tuned model converging to a lower pixel-level validation loss.

The results from our above experiments demonstrate that using a loss derived from high-level spectrogram patterns to tune the model does indeed improve performance over using only a pixel-level loss, for the \textit{vocals} and \textit{drums} sources, by 0.3 dB and 0.2 dB respectively (for the multi-channel model) over the samples in our study. While in itself, this is a valuable result and an improvement over the baseline model, it also lays down the case for further exploration of loss functions appropriate for music (or more generally, audio) data.

\section{Conclusion and Future Work}\label{future-work}

In this paper, we have demonstrated how using a high-level spectrogram feature loss, in addition to the standard pixel-level loss, can improve performance of a machine learning-based music source separation system. We believe that this is an improvement that could be generalized to related systems dealing with audio data. One area of improvement to the current work could be to explore spectrogram feature losses more customized to the audio/music domain. For eg., an audio classifier could be built and used in place of the VGG net. For the current application, it could be (for example) a network for discriminating between different musical instrument sounds. Secondly, to study the generalizability of our observation within deep learning-based music source separation, we could explore implementing alternative models described in literature for this task, with spectrogram feature losses (or their analog, for models that process the audio as a 1D signal).

\end{document}